\documentclass[twocolumn,prb]{revtex4}
\usepackage{amssymb}
\usepackage{graphicx}
\usepackage{dcolumn}
\usepackage{bm}

\begin{document}
\title[]{The electronic structure of a graphene quantum dot: Electric-field-induced evolution in two subspaces}
\author{Qing-Rui Dong}
\address{College of Physics and Electronics, Shandong Normal University, Jinan, Shandong, 250014, People's Republic of China}

\begin{abstract}
The tight-binding method is employed to investigate the effects of three typical in-plane electric fields on the electronic structure of a triangular zigzag graphene quantum dot.
The calculation shows that the single-electron eigenstates evolute independently in two subspaces no matter how the electric fields change.
The electric field with fixed-geometry gates chooses several scattered parts of the zero-energy eigenspace as the new zero-energy eigenstates, regardless of the field strength.
Moreover, the new zero-energy eigenstates remain unchanged and the associated levels are linear as the field strength.
In contrast, the new nonzero-energy eigenstates mix mutually and the associated levels are nonlinear as the field strength.
By comparing the effects of three electric fields, we demonstrate that the degeneracy of the zero-energy eigenstates accounts for the linearity of the associated levels.
\end{abstract}

\maketitle

\section{Introduction}
Graphene has attracted enormous interest both in theory and in experiments, due to its exceptional electronic properties\cite{Novoselov04} and great application potential in next-generation electronics.\cite{Wu11}
However, a gap has to be induced in the gapless graphene for its real applications in electronic devices.\cite{Ohta06,Zhou07}
For this purpose, graphene quantum dots have been proposed as one of the most promising kinds of graphene nanostructures.\cite{Ritter09}
With recent developments of fabrication techniques, it is possible to cut accurately the bulk graphene into different sizes and shapes, such as hexagonal zigzag quantum dots, hexagonal armchair quantum dots, triangular zigzag quantum dots and triangular armchair quantum dots.\cite{Zarenia11}

The electronic and magnetic properties of graphene quantum dots depend strongly on their shapes and edges.\cite{Ezawa07,Gu09,Potasz12}
Moreover, for zigzag graphene quantum dots, especially triangular dots, there appears a shell of degenerate states at the Dirac points and the degeneracy is proportional to the edge size.
The unique property of triangular zigzag quantum dots makes them potential components of superstructures acting as single-molecule spintronic devices.\cite{Wang09}
The electronic structure and total spin of triangular zigzag quantum dots can be tuned by changing a uniform electric field.\cite{Chen10,Ma12PRB}
The non-uniform electric fields can provide an equal electrostatic potential for the edges of triangular zigzag quantum dots, which allows the electrical linear control of the low-energy states.\cite{Dong13}
The magnetization of triangular graphene quantum dots with zigzag edges also can be manipulated optically.\cite{Gu13}
In particular, the electrical manipulation of the degenerate zero-energy states of such graphene quantum dots is quite important for the operation of related spintronic devices, since it is easier to generate the potential field through local gate electrodes than the optical or magnetic field.\cite{Zarenia11}
So, it is interesting to understand comprehensively the electric-field-induced evolution of the electronic structure in graphene quantum dots.

In this paper, we investigated the effects of three typical in-plane electric fields on the low-energy electronic structures of a triangular zigzag graphene quantum dot.
The calculations are mainly based on the tight-binding Hamiltonian with the nearest-neighbor approximation, which proves to give the same accuracy in the low-energy range as first-principle calculations.\cite{Abergel10}
Our result shows that the single-electron eigenstates evolute independently in two subspaces no matter how the electric fields change, which may be useful for the application of graphene quantum dots to electronic and photovoltaic devices.

\section{The tight binding model}

The low-energy electrical structure of a graphene quantum dot subjected to an in-plane electric field can be calculated by means of the tight-binding method.
The Hamiltonian equation of the system is $H|\Phi(\textbf{r})\rangle=E|\Phi(\textbf{r})\rangle$ and the tight-binding Hamiltonian with the nearest-neighbor approximation is\cite{Ma12,Chen10}
\begin{equation}\label{eqn:1}
H=\sum_{n}{(\varepsilon_{n}+U_{n})C^{+}_{n}C_{n}}+\sum_{<n,m>}{t_{n,m}C^{+}_{n}C_{m}},
\end{equation}
where $n$, $m$ denote the sites of carbon atoms in graphene, $\varepsilon_{n}$ is the on-site energy of the site $n$, $U_{n}$ is the electrostatic potential of the site $n$ (the electrostatic potentials applied to the whole quantum dot can be obtained by solving a Laplace equation), $t_{n,m}$ is the hopping energy and $C^{+}_{n}$ ($C_{n}$) is the creation (annihilation) operator of an electron at the site $n$. The summation $<n,m>$ is taken over all nearest neighboring sites. The effect of the electric field is to add the electrostatic potential $U_{n}$ to the on-site energy $\varepsilon_{n}$.
Due to the homogeneous geometrical configuration, the on-site energies and the hopping energies may be taken
as $\varepsilon_{n}=\varepsilon_{F}$ = 0 and $t_{n,m}=t$ = 2.7 eV.

The tight-binding Bloch function can be expressed as a linear superposition
\begin{equation}\label{eqn:0}
|\Phi(\textbf{r})\rangle=\sum_{n}c_{n}|\phi(\textbf{r}-\textbf{r}_{n})\rangle,
\end{equation}
where $\phi(\textbf{r}-\textbf{r}_{n})$ is the normalized $2p_{z}$ wave function for an isolated atom at the site $n$ and $c_{n}$ is the combination coefficient.
The matrix form of the tight-binding Hamiltonian can be obtained easily in the Wannier representation $|\phi(\textbf{r}-\textbf{r}_{n})\rangle$ and the low-energy spectrum of the graphene quantum dot can be calculated by diagonalizing the matrix.

Usually an electric field is generated by the gates with a fixed geometry and hence $U_{n}$ is proportional to the gate voltage (or the voltage difference) $U$:
\begin{equation}\label{eqn:2}
U_{n}=kUX_{n},
\end{equation}
where $k$ is a constant and $X_{n}$ is a function only dependent on $n$.
Despite the influence of an electric field, some eigenstates may remain unchanged and hence
\begin{eqnarray}\label{eqn:3}
& &\langle\Phi(\textbf{r})|H|\Phi(\textbf{r})\rangle\nonumber\\
&=&\langle\Phi(\textbf{r})|\sum_{n}{U_{n}C^{+}_{n}C_{n}}|\Phi(\textbf{r})\rangle\nonumber\\
& &+\langle\Phi(\textbf{r})|\sum_{<n,m>}{t_{n,m}C^{+}_{n}C_{m}}|\Phi(\textbf{r})\rangle\nonumber\\
&=&\sum_{n}[c^{*}_{n}\phi^{*}(\textbf{r}-\textbf{r}_{n})]\sum_{n}{(kUX_{n}C^{+}_{n}C_{n})}\sum_{n}[c_{n}\phi(\textbf{r}-\textbf{r}_{n})]\nonumber\\
& &+E\nonumber\\
&=&kUA+E
\end{eqnarray}
where $A=\sum_{n}{|c_{n}|^{2}X_{n}}$ is a constant and $E$ is the associated level at $U=0$.
Eq. (\ref{eqn:3}) means that the associated level is linear as $U$ when the eigenstate remains unchanged.
If the level is not degenerate, the converse is also true: the eigenstate remains unchanged when the associated level is linear as $U$.
If the level is degenerate, the linearity of the level means the associated eigenspace is unchanged.
So, if the level is not linear as $U$, the associated eigenstate or eigenspace would change.

\section{The electric fields and the low-energy electronic structures}

\begin{figure}[htp]
\includegraphics[height=12cm]{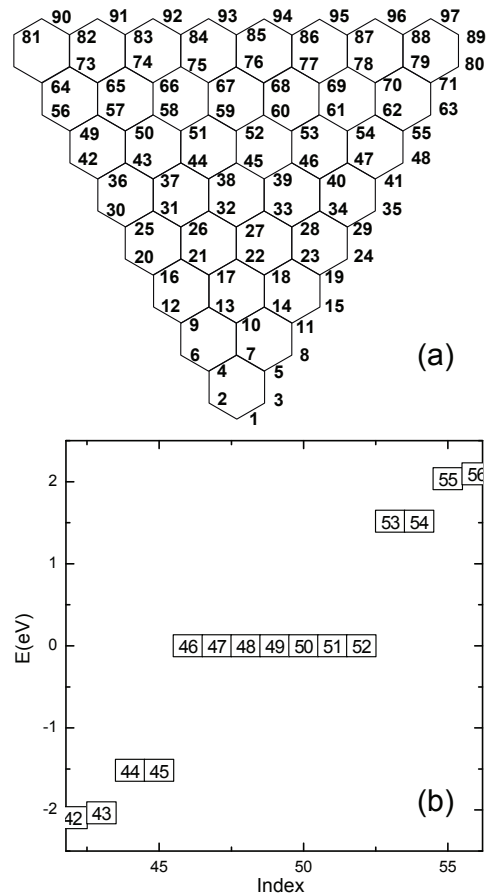}
\caption{\label{fig:1} (a) The geometrical structure of a triangular zigzag graphene quantum dot with the size $N_{s}$ = 8, where $N_{s}$ is the number of carbon atoms in each side of the quantum dot. The graphene quantum dot is labelled with the site $n$, which help to show the electron density later. (b) The low-energy spectrum of the graphene quantum dot as a function of the eigenstate index in the absence of an electric field.}
\end{figure}

The geometrical structure of a triangular zigzag graphene quantum dot is shown in Fig. \ref{fig:1} (a).
The number of carbon atoms in each side of the quantum dot is $N_{s}=8$.
The low-energy spectra of the graphene quantum dot in the absence of an electric field is shown in Fig. \ref{fig:1} (b), where the lowest fifteen eigenstates are presented and numbered from (42) to (56).
The seven orthonormal zero-energy eigenstates (46-52) are degenerate and span a 7-dimensional eigenspace denoted by $V_{1}$.
Other nonzero-energy eigenstates span the orthogonal complement space denoted by $V_{2}$.
The nonzero-energy orthonormal eigenstate (44, 45) as well as the eigenstates (53, 54) are degenerate and span respectively a two-dimensional eigenspace in $V_{2}$ .

\subsection{In a non-uniform electric field}

\begin{figure}[htp]
\includegraphics[height=5cm]{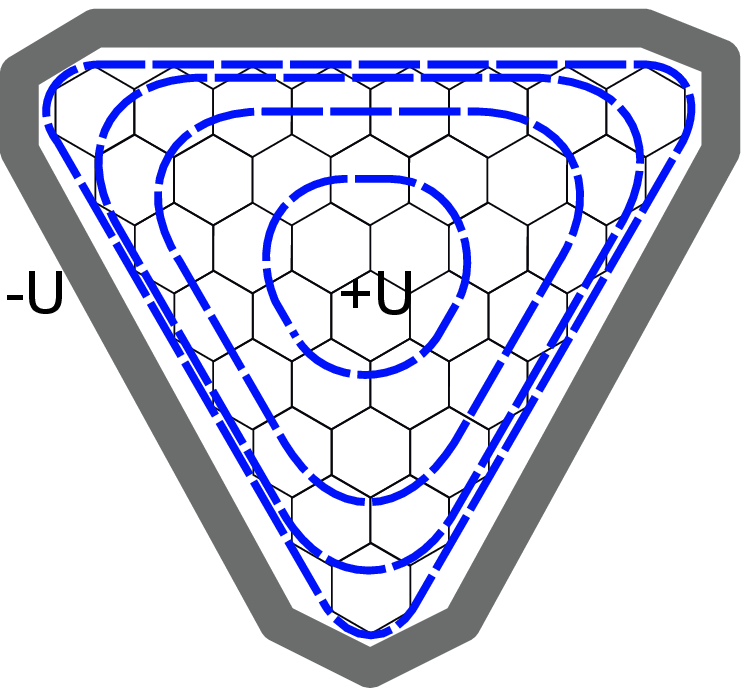}
\caption{\label{fig:EF11} The non-uniform electric field with a $C3$ rotation symmetry applied to a triangular zigzag graphene quantum dot ($N_{s}$ = 8). Two gates with electrostatic potentials $\pm U$ are applied outside and bottom of the quantum dot. The contour of the electrostatic potential is shown (blue dashed curves).}
\end{figure}

The electric field shown in Fig. \ref{fig:EF11} possesses the same $C3$ rotation symmetry as the graphene quantum dot.
Moreover the electric field can provide an equal electrostatic potential for all edge atoms, which is considered to accounted for the electrical linear control of the zero-energy states.\cite{Dong13}
The designed gates work in a similar way as a lateral gated quantum dot is created at a semiconductor heterojunction containing a two-dimensional electron gas.

\begin{figure}[htp]
\includegraphics[height=6cm]{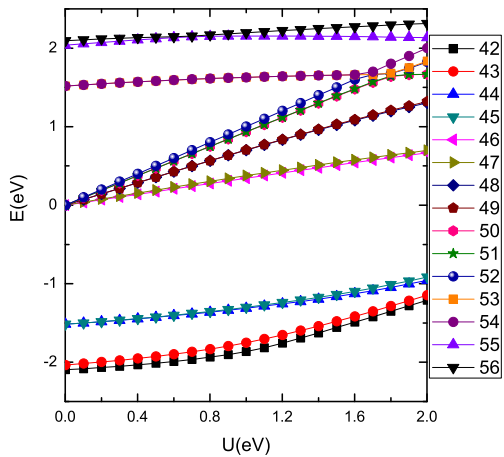}
\caption{\label{fig:EF12} The low-energy spectrum of the graphene quantum dot ($N_{s}$ = 8) subjected to the non-uniform electric field shown in Fig. \ref{fig:EF11}. The eigenstate indexes correspond to those in Fig. \ref{fig:1} (b). }
\end{figure}

\begin{figure}[htp]
\includegraphics[height=11cm]{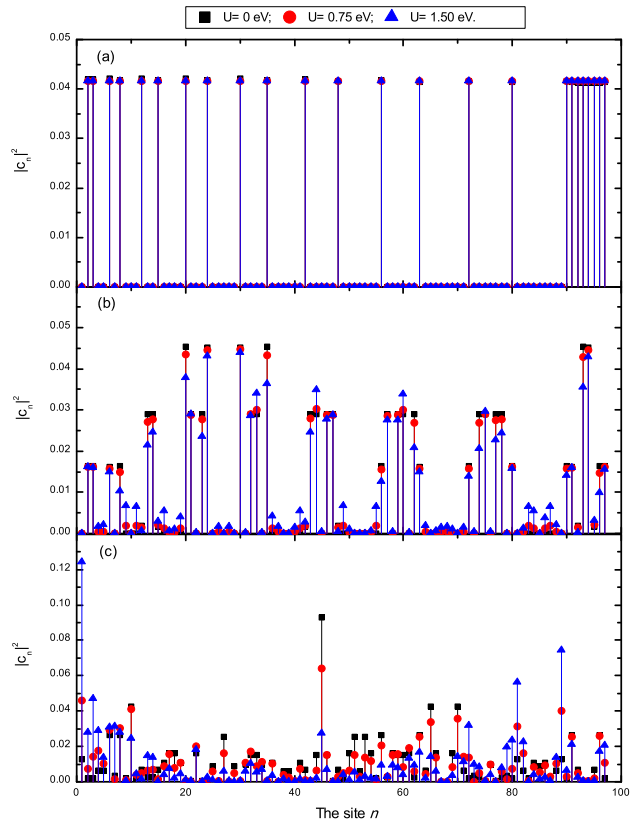}
\caption{\label{fig:EF13} The density of three nondegenerate eigenstates. (a)-(c) correspond respectively to the eigenstate (52), (46) and (43) shown in Fig. \ref{fig:EF12}.}
\end{figure}

Fig. \ref{fig:EF12} shows the low-energy spectrum of the graphene quantum dot ($N_{s}$ = 8) subjected to the electric field.
Since the electrostatic potential does not possess the translational symmetry, the zero-energy eigenspace changes from the 7-dimensional subspace $V_{1}$ into several scattered parts, including one nondegenerate eigenstate (52), two two-dimensional eigenspace (50,51)/(48,49) and one quasi-degenerate eigenspace (46,47).
As increasing $U$, the seven levels vary linearly, which implies that the associated eigenstate or eigenspaces do not change significantly according to Eq. (\ref{eqn:3}).
To a nondegenerate eigenstate, the stability of the eigenstate can be shown by the corresponding probability density.
Fig. \ref{fig:EF13}(a) shows the probability density of the eigenstate (52), which indicates that the eigenstate remains unchanged.
Obviously, the electric field, or rather the gate geometry, chooses several scattered parts of the subspace $V_{1}$ as the new zero-energy eigenstates and then these scattered parts remain unchanged, regardless of the field strength.
Hence, the zero-energy eigenstates can be considered to always evolute in $V_{1}$ as increasing $U$.
The level of the quasi-degenerate eigenspace (46,47) remains linear on the whole, which implies that the quasi-degenerate eigenspace does not change significantly.
If the energy difference between the eigenstate (46) and (47) can not be neglected, the degeneracy disappears and the linearity is not perfect.
The imperfect linearity implies that the eigenstates (46) and (47) change lightly.
The probability density of the eigenstate (46) changes lightly (see Fig. \ref{fig:EF13}(b)) and the characteristic can also be seen in the probability density of the eigenstate (47).
According to the orthogonality of the eigenstates, the eigenstates (46) and (47) can be considered to interact lightly since other zero-energy eigenstates remain unchanged.

In contrast, the nonzero-energy levels generally are nonlinear as increasing $U$, which implies that the associated eigenstates or eigenspaces change.
As a typical example, the probability density of the eigenstate ($43$) shows that the eigenstate changes significantly as increasing $U$ (see Fig. \ref{fig:EF13}(c)).
According to the orthogonality of the eigenstates, the eigenstates in $V_{1}$ are perpendicular to the subspace $V_{2}$ and remain unchanged, which implies that the nonzero-energy eigenstates can be considered to always evolute in $V_{2}$ as increasing $U$.
The conclusion can also be proved by the fact that there is not a distinct anticrossing between the levels associated to $V_{1}$ and $V_{2}$ (see Fig. \ref{fig:EF12}).
According to the completeness of the eigenstates, the eigenstates in $V_{2}$ can be considered to mix mutually as $U$ increases.

As $U$ increases, the eigenstates in $V_{1}$ remain unchanged while the eigenstates in $V_{2}$ mix mutually.
By taking into account the difference, one can assume that the degeneracy of the zero-energy eigenstates accounts for the linearity of the levels.
Moreover, it will be shown in the following section that all two-dimensional eigenspaces disappear when the electric field loses the $C3$ symmetry.

\subsection{In a uniform electric field}
\begin{figure}[htp]
\includegraphics[height=5cm]{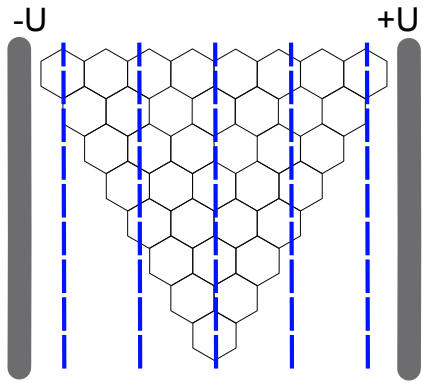}
\caption{\label{fig:EF21} The uniform electric field applied to a triangular zigzag graphene quantum dot ($N_{s}$ = 8). Two gates with electrostatic potentials $\pm U$ are applied to the left and right of the quantum dot. The contour of the electrostatic potential is shown (blue dashed curves).}
\end{figure}

\begin{figure}[htp]
\includegraphics[height=6cm]{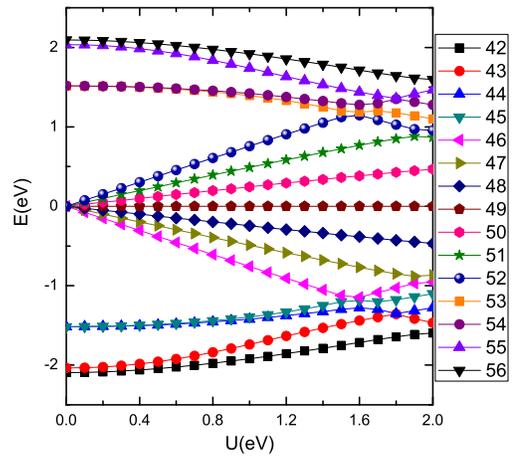}
\caption{\label{fig:EF22} The low-energy spectrum of the graphene quantum dots ($N_{s}$ = 8) subjected to the uniform electric field shown in Fig. \ref{fig:EF21}. The eigenstate indexes correspond to those in Fig. \ref{fig:1} (b).}
\end{figure}

If it is true that the degeneracy of the zero-energy eigenstates accounts for the linearity of the levels, a uniform electric field, ever considered not to lead to the linearity\cite{Ma12PRB}, also can do this.
In order to prove the viewpoint, a uniform electric field is presented in Fig. \ref{fig:EF21}, which does not possess the $C3$ symmetry and can not provide an equal potential for all edge atoms.
Moreover, the low-energy spectrum of the graphene quantum dot subjected to the electric field is shown in Fig. \ref{fig:EF22}.

\begin{figure}[htp]
\includegraphics[height=7cm]{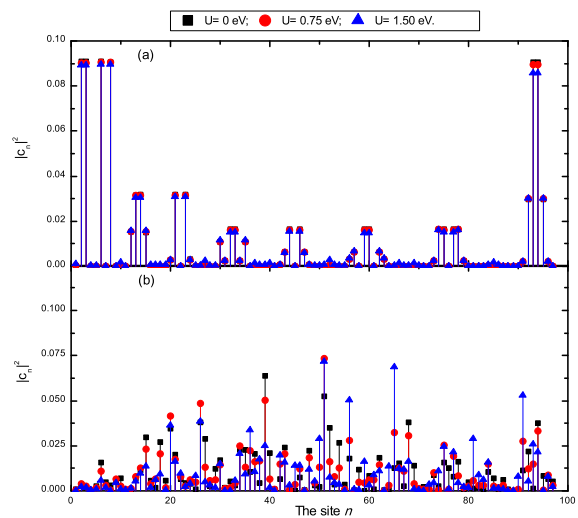}
\caption{\label{fig:EF23}  The density of two nondegenerate eigenstates. (a) and (b) correspond respectively to the eigenstate (49) and (53) in Fig. \ref{fig:EF22}.}
\end{figure}

Since the electric field leads to a lower level of symmetry than the electric field shown in Fig. \ref{fig:EF11}, the zero-energy eigenspace changes from the 7-dimensional subspace $V_{1}$ into seven nondegenerate eigenstates.
As $U$ increase, the seven associated levels vary linearly, which implies that each zero-energy eigenstate does not change significantly.
As a typical example, the probability density of the eigenstate ($49$) is shown in Fig. \ref{fig:EF23}(a), which indicates that the eigenstate remains unchanged.
The stability also can be seen in the probability density of other six eigenstates.
Obviously, the seven nondegenerate eigenstates are chosen from the 7-dimensional subspace $V_{1}$ by the gate geometry and then remain unchanged, regardless of the field strength.
This also prove that the degeneracy of the zero-energy eigenstates should account for the linearity of the levels.
Without two-dimensional eigenspaces due to the symmetry of the electric field, the interaction between quasi-degenerate eigenstates also does not occur and hence the linearity is more perfect.

All nonzero-energy levels are nonlinear, which implies that the associated eigenstates change.
As a typical example, the probability density of the eigenstate (53) indicates that the eigenstate changes significantly as increasing $U$ (see Fig. \ref{fig:EF23}(b)).
According to the previous analysis, the nonzero-energy eigenstates mix mutually in $V_{2}$ as $U$ increases.

\subsection{In an electric field with random potential distribution}

Since the degeneracy of the zero-energy eigenstates accounts for the linearity of the levels, one can make some predictions on the electric field with arbitrary fixed-geometry gates.
The electric field should choose seven nondegenerate eigenstates as the new zero-energy eigenstates according to the gate geometry if the arbitrary electric field possesses a lower level of symmetry.
Moreover, the electric field, as well as the two electric fields mentioned previously, should keep the new zero-energy eigenstates unchanged in $V_{1}$ while the new nonzero-energy eigenstates mix mutually in $V_{2}$.

\begin{figure}[htp]
\includegraphics[height=5cm]{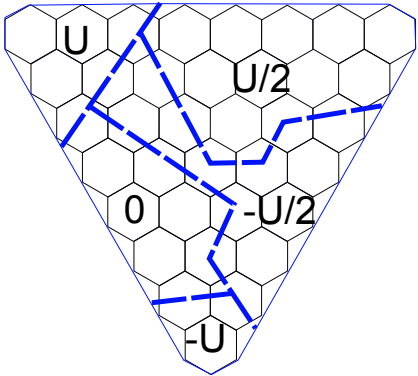}
\caption{\label{fig:EF31} The imaginary electric field with a random potential distribution applied to a triangular zigzag graphene quantum dot ($N_{s}$ = 8). The contour of the electrostatic potential is shown (blue dashed curves).}
\end{figure}

\begin{figure}[htp]
\includegraphics[height=6cm]{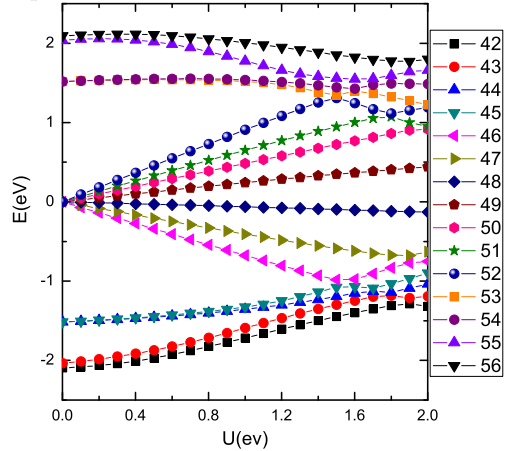}
\caption{\label{fig:EF32} The low-energy spectrum of the graphene quantum dots ($N_{s}$ = 8) subjected to the imaginary electric field shown in Fig. \ref{fig:EF31}. The eigenstate indexes correspond to those in Fig. \ref{fig:1} (b).}
\end{figure}

In order to verify these predictions, an imaginary electric field is presented in Fig. \ref{fig:EF31}, which receives randomly an imaginary potential distribution.
The low-energy spectrum of a triangular zigzag graphene quantum dot ($N_{s}$ = 8) subjected to the electric field is shown in Fig. \ref{fig:EF32}.
As $U$ increases, the levels of the seven zero-energy eigenstates vary linearly and all levels of the nonzero-energy eigenstates vary nonlinearly, which implies that the effect of the electric field agrees with the above predictions.
Moreover, the zero-energy levels and the spaces between the levels are dependent on the random potential, which implies that it is more effective to modulate the zero-energy eigenstates by changing the gate geometry.

\subsection{In a arbitrarily changing electric field}

If the gate geometry changes, for example, from the electric field in Fig. \ref{fig:EF11} to the electric field in Fig. \ref{fig:EF21} and then to the electric field in Fig. \ref{fig:EF31}, the zero-energy eigenstates will also change.
However, the evolution of the zero-energy eigenstates is confined in $V_{1}$ since the zero-energy eigenstates for any gate geometry are chosen from $V_{1}$ according to the previous analysis.
Moreover, the evolution of the nonzero-energy eigenstates is confined in $V_{2}$ according to the orthogonality of the eigenstates.
That is to say, no matter how the gate geometry and voltage change, the eigenstates evolute independently in two subspace $V_{1}$ and $V_{2}$.

\section{Summary}
In summary, we investigated the effects of three typical in-plane electric fields on the electronic structure of a triangular zigzag graphene quantum dot.
The results show that no matter how the electric fields change, the single-electron eigenstates evolute independently in two subspaces $V_{1}$ and $V_{2}$.
The electric field with fixed-geometry gates chooses several scattered parts of the subspace $V_{1}$ as the new zero-energy eigenstates.
Moreover, the eigenstates in $V_{1}$ remain unchanged and the associated levels are linear as $U$ due to the degeneracy of the zero-energy eigenstates.
In contrast, the eigenstates in $V_{2}$ mix mutually and the associated levels are nonlinear as $U$.
Two-dimensional eigenspaces can be removed by lowering the symmetry level of the electric field, which helps to keep the zero-energy eigenstates unchanged and to keep the associated levels linear as the field strength.
The calculation implies that it is more effective to modulate the zero-energy eigenstates by changing the gate geometry.
Our results provide insight into the electric-field-induced evolution of the electronic states in a graphene quantum dot and may be useful for the application of graphene quantum dots to electronic and photovoltaic devices.

\end{document}